\documentclass[showpacs,twocolumn]{revtex4-1}
\usepackage[utf8]{inputenc}
\usepackage{textcomp}
\usepackage{amsfonts,amsmath,amssymb,graphicx,dsfont}
\newcommand\tr{{\rm Tr}\;}
\newcommand\ket[1]{\,| #1 \rangle}
\newcommand\bra[1]{\langle #1 |\,}
\newcommand\proj[2]{\ket{#1}\!\bra{#2}}
\newcommand\media[1]{\left\langle #1 \right\rangle}

\newcommand\del[2]{\frac{\partial #1}{\partial #2}}
\newcommand\Zeta{\mathcal{Z}}

\begin{document}

\title{Thermodynamic cost of acquiring information}

\author{Kaonan Micadei}
\email{kaonan.bueno@ufabc.edu.br}
\affiliation{Centro de Ci\^{e}ncias Naturais e Humanas, Universidade Federal do ABC, R.  Santa Ad\'{e}lia 166, 09210-170 Santo Andr\'{e}, S\~{a}o Paulo, Brazil}

\author{Roberto M. Serra}
\email{serra@ufabc.edu.br}
\affiliation{Centro de Ci\^{e}ncias Naturais e Humanas, Universidade Federal do ABC, R.  Santa Ad\'{e}lia 166, 09210-170 Santo Andr\'{e}, S\~{a}o Paulo, Brazil}

\author{Lucas C. C\'{e}leri}
\email{lucas@chibebe.org}
\affiliation{Instituto de F\'{i}sica, Universidade Federal de Goi\'{a}s, 74.001-970, Goi\^{a}nia, Goi\'{a}s, Brazil}

\begin{abstract}
Connections between information theory and thermodynamics have proven to be very useful to establish bounding limits for physical processes. Ideas such as Landauer's erasure principle and information assisted work extraction have greatly contributed not only to enlarge our understanding about the fundamental limits imposed by nature, but also to enlighten the path for practical implementations of information processing devices. The intricate information-thermodynamics relation also entails a fundamental limit on parameter estimation, establishing a thermodynamic cost for information acquisition. We show that the amount of information that can be encoded in a physical system by means of a unitary process is limited by the dissipated work during the implementation of the process. This includes a thermodynamic trade-off for information acquisition. Likewise, the information acquisition process is ultimately limited by the second law of thermodynamics. This trade-off for information acquisition may find applications in several areas of knowledge.
\end{abstract}

\maketitle

Information theory first met thermodynamics when Maxwell introduced his famous Demon \cite{Maxwell}. This relation became clear with Brillouin's treatment of the information entropy (due to Shannon) and the thermodynamic entropy (due to Boltzmann) on the same footing \cite{Brillouin}. Many advances linking these two apparently distinct areas have been achieved since then, with one of the most remarkable being ascribed to Landauer's erasure principle \cite{Landauer}. This principle, introduced as an effectively way to \textit{exorcize} Maxwell's Demon, states that erasure of information is a logically irreversible process that must dissipate energy. More recently, developments in this directions include theoretical and experimental investigations of Landauer's principle and its consequences \cite{Vlatko, Berut}, work extraction by feedback control of microscopic systems \cite{Ueda,Ueda1,Ueda2,Ueda3,Ueda4}, and links between the second law of thermodynamics and two fundamental quantum mechanical principles, i.e., the wave-function collapse \cite{Hormoz} and the uncertainty relation \cite{Hanggi}. Here, we introduce a thermodynamic trade-off for information acquisition, which relates the uncertainty of the information acquired in a parameter estimation process with the dissipated work by the encoding process. This trade-off relation is obtained by a formal connection between an elusive quantity from estimation theory, named Fisher information \cite{Fisher,Fisher1,Kullback,Hayashi}, and the Jarzynski equality \cite{Jarzynski}.

\section{Results}     

\begin{figure}
\includegraphics[scale=0.32]{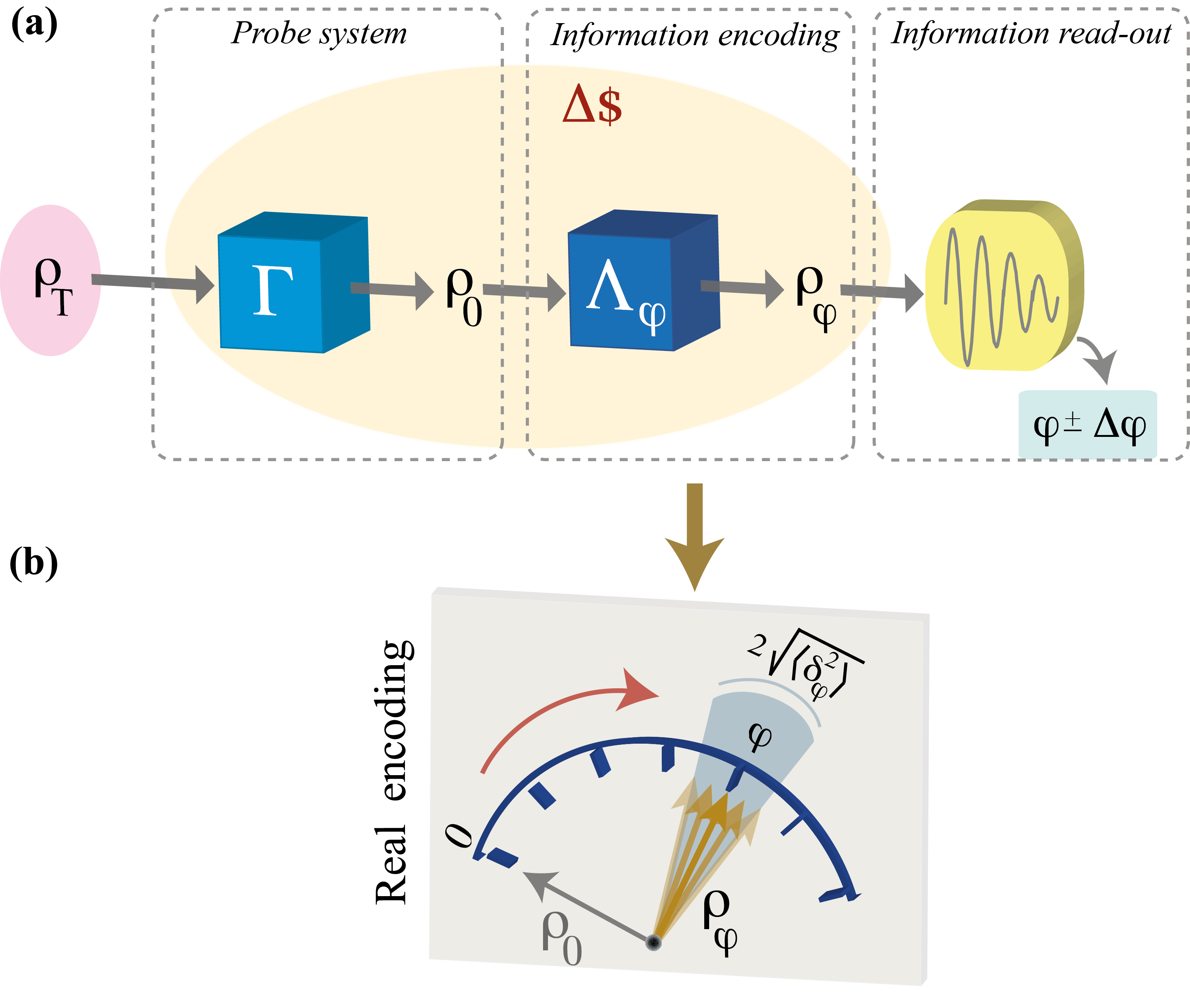}
\caption{General sketch of a parameter estimation process. (a) The estimation process works as follows: A probe system initially in the thermal equilibrium, $\rho_{T}$, at a given temperature $T$, is prepared in a suitable blank state $\rho_{0}$ (through a process $\Gamma$). Then, $\Lambda_{\varphi}$ acts on the probe in order to encode some information about the desired parameter $\varphi$. The thermodynamic cost of the probe state preparation and encoding of information is $\Delta\textdollar$. The final step is the read-out of the encoded probe $\rho_{\varphi}$. This measurement process results in a probability distribution, $p_{\varphi}$, that contains the information about $\varphi$. After some statistical manipulation of this measured distribution, an estimation for the value of $\varphi$ is obtained with the mean square root error $\Delta\varphi$. (b) Pictorial representation of an imperfect encoding of information. The probe state is represented as a pointer and the encoding process $\Lambda_{\varphi}$ has a (minimum) finite precision $\delta_{\varphi}$.}
\end{figure}

Natural sciences are based on experimental and phenomenological facts. Parameter estimation protocols have a central role to the observation of new phenomena or to validate some theoretical prediction. Suppose, we want to determine the value of some parameter, let us say $\varphi$. This task can be accomplished, generally, by employing a probe, $\rho_{T}$. We will assume that the probe state is initially in thermal equilibrium at absolute temperature $T$, so $\rho_{T}$ is the canonical equilibrium (Gibbs) state \cite{Popescu}. In order to extract some information about the parameter $\varphi$, the probe could be prepared (through a process $\Gamma$) in a suitable \textit{blank} state represented by $\rho_{0}$. Then the probe is transformed by a unitary process $\Lambda_{\varphi}$ in order to encode information about the parameter on the probe state $\rho_{\varphi}$. In general, in real world applications, these operations (probe state preparation and encoding of information) are logically irreversible and therefore, must have an energetic cost. The effectiveness of the estimation (metrology) process depends on how information is encoded in the probe system. This encoding operation consumes some work from a thermodynamic point of view. An estimation of the parameter $\varphi$ can be obtained by a suitable read-out of the encoded probe system $\rho_{\varphi}$. The aforementioned protocol (and also outlined in Fig. 1(a)) abstractly summarizes the operation of almost all high-precision measurement devices. Employing this general framework, we show that the uncertainty (the mean square root error) $\Delta\varphi$ of an estimation process is limited by a general physical principle
\begin{equation}
\Delta\$\cdot\Delta\mathcal{I}_{\varphi} \geq \frac{k_{B}}{2},
\label{main}
\end{equation} 
where $k_{B}$ is Boltzmann's constant and the thermodynamic trade-off for information acquisition is defined as the mean dissipated work $\langle\mathcal{W}_{D}\rangle$ at a given temperature $T$ as $\Delta\textdollar = \langle\mathcal{W}_{D}\rangle / T$ and the relative acquired information as $\Delta\mathcal{I}_{\varphi} = (\Delta\varphi)^{2} / \delta^{2}_{\varphi}$. $\delta_{\varphi}$ is a quantity describing the accuracy of the encoding process. Roughly speaking, $\delta_{\varphi}$ is the precision of the experimental device used to implement $\Lambda_{\varphi}$ (the minimum scale for $\varphi$, see Fig. 1(b) and the Supplementary Information). The symbol $\langle\cdot\cdot\cdot\rangle$ represents the mean value with respect to an ensemble of measurements. The physical quantities appearing in Eq. (\ref{main}) are highly process dependent and must be carefully defined in each physical set-up. We proceed by analysing the physical meaning of Eq. (\ref{main}) and discussing some of its implications, postponing its derivation. 

The work consumed in the parameter estimation process could ultimately be attributed to the logical irreversibility of information encoding. The first step for any estimation protocol is to prepare the probe. If we employ an out of equilibrium probe, we have to erase the thermal state to prepare the probe in a suitable blank state and it has some energetic cost  (from the Landauer's principle). The second step, i.e., the encoding of information in the probe state ($\Lambda_{\varphi}$), in a realistic apparatus is not perfect and must therefore also dissipate some energy. In order words, the finite precision $\delta_{\varphi}$ of the encoding operation implies logical irreversibility and, as a consequence, work dissipation. For the sake of clarity, let us discuss these issues in two physical contexts. 

\begin{figure}
\includegraphics[scale=0.18]{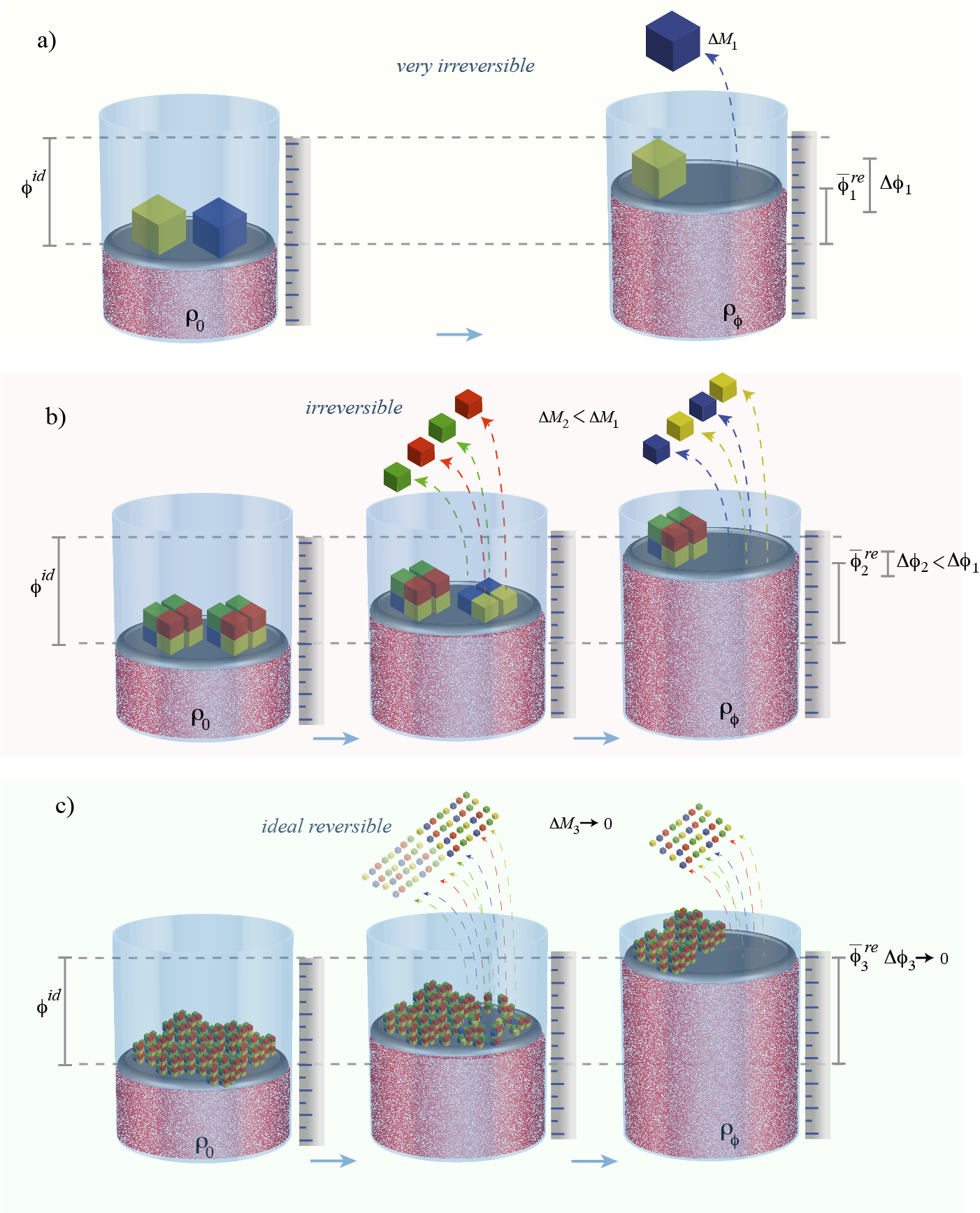}
\caption{Illustration of a parameter estimation process employing a classical apparatus. The apparatus itself is composed of a gas confined within a cylindrical chamber with a movable piston. Information is encoded in the piston position adding or removing some amount of mass above the piston. (a) A very irreversible process where the information encoding is performed  moving suddenly a large amount of mass in a single shot. (b) In this case the information encoding is performed moving small portions of mass, this still an irreversible process which dissipating less work than the first one. (c) Sketch of an idealized reversible process where the amount of mass above the piston is removed in an adiabatic way.}
\end{figure}

First, we consider a simple classical thermodynamic system. In this classical setting all the quantities appearing in Eq. (\ref{main}) are naturally defined. Nevertheless, our results can be applied to both (out of equilibrium) classical and quantum systems. Let us suppose that our apparatus is composed of a gas confined within a cylindrical chamber in which the upper base is made of a movable piston with some amount of matter ($M$) placed over it as sketched in Fig. 2. The gas is our probe system and the position of the movable piston indicates the probe state. Defining $\varphi_{0}$ as the equilibrium position of the piston, we can encode information on this system introducing (or removing) some amount of matter over the piston. The information is encoded by the displacement $\varphi$ of the piston from its initial equilibrium position $\varphi_{0}$. Considering the parameter estimation processes as described in Fig. 1(a), we have the following steps: (i) Initially, the system is in thermal equilibrium at temperature $T$, being described by the state $\rho_{T}$, corresponding to piston position $\varphi_{0}$ (with some amount of mass $M$ on the piston). In this situation, all the forces acting on the piston are in equilibrium and its position is fixed (except for thermal fluctuations). In this context, the equilibrium state $\rho_{T}$ could be a suitable probe ($\rho_{0}=\rho_{T}$), so the probe preparation process ($\Gamma$ of Fig. 1(a)) does not dissipate work. (ii) By removing (in an adiabatic or in a non-adiabatic way) some amount of mass ($\Delta M$) on the piston the information can be encoded into the probe. Due to the unbalance of forces, this operation drives the piston to a new equilibrium position described by the state $\rho_{\varphi}$, corresponding to position $\varphi$, thus encoding the information into the probe state. During such implementation a certain amount of work must be employed and part of it may be dissipated into the environment. In order to get a good estimation of the encoded information, this protocol should be repeated $N$ times or $N$ identical copies of such a system should be employed. Each realization of the protocol is driven by an amount of performed work $w^{i}$. The mean applied work is then given by $\langle\mathcal{W}\rangle = N^{-1}\sum p(w^{i})w^{i}$, with $p(w^{i})$ being the measured work probability distribution for the whole ensemble of realizations. (iii) Measuring the piston position in all realizations, we obtain an estimation $\varphi = \langle \varphi\rangle \pm \Delta \varphi$ for the parameter, where $\langle \varphi\rangle = N^{-1}\sum p(\varphi_{i})\varphi_{i}$ is the mean value of the piston displacement, $p(\varphi_{i})$ is the observed probability distribution and $\Delta \varphi$ is the mean square root deviation.

In the example explored above the work is dissipated in step (ii). When we remove a certain amount $\Delta M$ of mass from the top of the piston, the gas expands until a new equilibrium position is reached. The amount of removed mass determines how much the piston position changes and, ultimately, it will also determine how much work will be dissipated. If the whole mass is removed in just one shot (Fig. 2(a)), there will be a huge amount of dissipated work, since the gas will expand from the initial state (corresponding to $\rho_{T}$) to the final one (corresponding to $\rho_{\varphi}$), through a sudden and irreversible path \cite{thermo}. The amount of dissipated work is given by the second law of thermodynamics as $\langle\mathcal{W}_{D}\rangle = \langle\mathcal{W}\rangle - \Delta F$, where $\Delta F$ is the difference in free energy between the final and the initial states and $\langle\mathcal{W}\rangle$ is the mean invested work during the encoding process. In this case, the final position of the piston (the encoded information) will deviate from the predicted ideal reversible one. This example clearly shows that some information is lost in the encoding process due to work dissipation caused by finite changes in the system (irreversibility). 

To minimize the information loss and, consequently, to improve the precision of the protocol, we have to diminish dissipation as much as possible. This can be accomplished by removing the mass in small portions as depicted in Fig 2(b), with the limit being the idealized reversible process, for which $\langle\mathcal{W}_{D}\rangle = 0$ (Fig. 3(c)). In this case, Eq. (\ref{main}) seems to be flawed, but a deeper analysis reveals that this is not the case. For the implementation of a reversible process, we must take the limit $\Delta M\rightarrow 0$ (the process must be implemented in a quasi-static way). But, this limit implies $\delta_{\varphi}\rightarrow 0$, since $\delta_{\varphi}$ is the minimum step size that the piston is able to move, i.e., the minimum change in the system (see below and the Supplementary Information for formal details). In the limit of reversible processes we can read-out all the information encoded in the probe. On the other hand, in the real word the ``scale'' of the encoding apparatus is finite. In this case, the minimum amount of mass that can be removed is finite, this also introduces a minimum step size for the position of the piston, i.e., $\delta_{\varphi}>0$. This inevitably leads to information loss in the encoding process due to work dissipation. In fact, any realistic encoding apparatus with a finite precision (scale) is irreversible, therefore the apparatus must dissipate work, introducing uncertainty in the parameter estimation as ultimately bounded by Eq. (\ref{main}). Although we have explored a specific example, this statement is independent of the physical system, just like the fact that irreversible processes (associated with finite changes in the system) must increase entropy.   

Two other important limits are the zero and infinity temperature. Regarding, as discussed earlier, that all real processes are irreversible and information is inevitably lost, we have $\langle\mathcal{W}_{D}\rangle > 0$, implying that $\delta_{\varphi} >0$. Let us assume that both dissipated work $\langle\mathcal{W}_{D}\rangle$ and the encoding accuracy $\delta_{\varphi}$ are constants with respect to the temperature. When $T\rightarrow \infty$ Eq. (\ref{main}) leads to $(\Delta\varphi)^2 \rightarrow \infty$. The observer cannot obtain any information encoded by the process $\Lambda_\varphi$. Actually, we cannot encode any information in this limit due to the infinite amplitude of thermal fluctuations, which wash out all the information, no matter how precise the encode process is. In the opposite limit, $T\rightarrow 0$, we have $(\Delta\varphi)^{2} \geq 0$. For classical systems, this is a valid limit and the inequality could be, in principle, saturated. However, due to the third law of thermodynamics, it is not allowed for quantum systems to reach this limit. $(\Delta\varphi)^{2}$ is always greater than zero due to quantum fluctuations. 

The bound presented in Eq. (\ref{main}) also holds for quantum strategies for parameter estimation employing out of equilibrium probes. Now, let us consider a standard interferometric strategy to estimate a phase shift between two states. This task can be accomplished observing the probability for the measurement of the probe in a suitable basis. We are going to label the two states by $\left| 0 \right\rangle$ and $\left| 1 \right\rangle$. A suitable probe in this case is a balanced superposition, $\left| \psi_0 \right\rangle=\left(\left| 0 \right\rangle + \left| 1 \right\rangle\right)/\sqrt{2}$  ($\rho_{0}=\left| \psi_0 \right\rangle \left\langle \psi_0\right| $). The probe state is out of equilibrium and some work has to be dissipated to prepare it through the process $\Gamma$ depicted in Fig. 1(a). This process could be a suitable post-selected projective measurement on the thermal equilibrium state $\rho_{T}$. This operation erasures information and dissipates some energy according to Landauer's principle. The encoding of information could be employed by a phase shifter, as for example: $U(\varphi)=e^{i\phi \left| 1 \right\rangle \left\langle 1\right|}$. In a real interferometer the minimum step size $\delta_\varphi$ to encode the phase $\varphi$ on the probe state is finite. Therefore, we have an imperfect encode (as pictorially described in Fig. 1(b)) and in an ensemble of realizations, the evolution is irreversible. This finite accuracy of the phase shifter ($\delta_\varphi > 0$) implies information loss and work consumption in the encoding process $\Lambda_\varphi$. In fact, the probe state preparation is also a non-ideal process introducing another source of dissipated work. The bound for information-acquisition in the out of equilibrium quantum context is also given by Eq. (\ref{main}). 

\section{Discussions} 

We introduced a physical principle that bounds information acquisition, Eq. (\ref{main}), derived from an information-theoretic relation associated to Jarzynski equality \cite{Jarzynski} and from the Cram\'{e}r-Rao \cite{Cramer,Rao} relation. This is a general result, applicable to classical or quantum contexts, stating that the amount of information that can be encoded by means of a unitary process is limited by the dissipated work (due to logical irreversibility) during the implementation of the estimation process. This conclusion reveals a deep connection between Metrology and Thermodynamics, implying that the physical limit for the precision of a parameter estimation process (which is equivalent to encoding and decoding information processes) is given by Thermodynamics. Moreover, the lower bound on the uncertainty about the estimation of a given parameter is zero only in the thermodynamic limit of reversible (adiabatic) processes (imposed by the second law). 
 
The inequality (\ref{main}) could be conceived as a counterpart of Landauer's principle, as both of them are assertions about the work cost of information (acquisition or erase). Furthermore, it would be interesting to investigate the relation of the results herein with generalized uncertainty relations. At this point, it is reasonable to presume that the basic principles of quantum mechanics itself are probably subtly connected to the second law of thermodynamics \cite{Hanggi,Wehner,Oscar} in an informational scenario.  
 
From the point of view of the experimental verification of Eq. (1), it is important to precisely establish the system, in order to define all the quantities involved, such as the work employed in the process and how the information is encoded and read-out.
 
Discussing the fundamentals of physics, Planck has argued that the number of dimensional fundamental constants in nature should be equal to four \cite{Planck}: the Newtonian gravitational constant $G$, the speed of light $c$, Planck's and Boltzmann's constants $h$ and $k_{B}$, respectively. The authors of Ref. \cite{Vanzella} concluded that this number should be two, chosen between $G$, $c$ and $h$, having discarded Boltzmann's constant for being a conversion factor between temperature and energy. In Ref. \cite{Bohr}, the viewpoint that Planck's constant is superfluous was advocated and $k_{B}$ was also discarded for the same reason given in \cite{Vanzella}. If we define temperature as twice the mean value of the energy stored in each degree of freedom of a system in thermal equilibrium, $T = 2\langle E_{0}\rangle$, $k_{B}$ turns out to be a dimensionless quantity equal to one and Eq. (\ref{main}) becomes
\begin{equation}
\frac{\langle\mathcal{W}_{D}\rangle}{\langle E_{0}\rangle}\cdot\left(\frac{\Delta\varphi}{\delta_{\varphi}}\right)^{2} \geq 1,
\end{equation}
which means that the precision of the information acquired in a parameter estimation process is limited by the mean dissipated work per degree of freedom of the encoding system. On a more practical ground, inequality (\ref{main}) is quite meaningful for technological applications on metrology relating the reversibility of hight precision measurement device with its efficiency.  

\section{Methods} 

Here, we outline the derivation of Eq. (\ref{main}), postponing the details to the Supplementary Information. Consider again the general estimation process described in Fig. 1(a). To inspect how a given unbiased estimator for the parameter $\varphi$ is close to the real encoded information, we can use the so-called Cram\'{e}r-Rao bound \cite{Cramer,Rao}
\begin{equation}
(\Delta\varphi)^{2} \geq \frac{1}{\mathcal{F}},
\label{cramerrao}
\end{equation}
where $\mathcal{F}$ is the Fisher Information, usually defined as $ \mathcal{F} = \int dx p_{\varphi}\left(x\right)\left(\frac{\partial \ln p_{\varphi}\left(x\right)}{\partial\varphi}\right)^{2} $. $p_{\varphi}\left(x\right)$ is the probability distribution for the best read-out strategy of the encoded probe and it contains the information about $\varphi$. For our proposal, it will be interesting to express the Fisher information in terms of a relative entropy as \citep{Kullback,Hayashi}   
\begin{equation}
\mathcal{F} \approx 2 \frac{S(p_{\varphi}^{re}||p_{\varphi}^{id})}{\delta_{\varphi}^{2}},
\label{relativeent-fisher}
\end{equation}
with $p_{\varphi}^{re}$ being the read-out probability distribution obtained in a real (irreversible, non-ideal) experiment, $p_{\varphi}^{id}$ being the read-out probability distribution of an ideal (reversible) parameter estimation protocol and $\delta_{\varphi}$ is the accuracy of the process. Here, we refer to the ideal process as the limit process of the second law, i.e., the adiabatic process where all the work in converted into free energy ($\langle\mathcal{W}\rangle - \Delta F = 0$). By the real (irreversible) experiment we mean that the parameter estimation apparatus is non-ideal, in the sense that the process running in such apparatus will dissipate some amount of work ($\langle\mathcal{W}\rangle - \Delta F > 0$). In this case, we consider a slight non-ideal process working very near to the reversible limit, since we are interested in a high precision measurement apparatus. The approximation presented in Eq. (\ref{relativeent-fisher}) is a very good approximation in this setting (see the last section of the Supplementary Information for details).              

Next, we relate the Fisher information with information loss and the dissipated work through a formal relation between work and information obtained in the first section of the Supplementary Information. Considering that a system is driven, through the injection of work $\mathcal{W}$ by an external agent, from the initial equilibrium state to some final one, Jarzynski proved that \cite{Jarzynski}
\begin{equation}
\left\langle e^{-\mathcal{W}/k_{B}T}\right\rangle = e^{-\Delta F/k_{B}T},
\label{jar}
\end{equation}
where the mean is computed over the ensemble of realizations and $\Delta F$ is the free energy difference between the final and the initial system's states. $T$ is the temperature of the initial equilibrium state. In the Supplementary Information we obtain the following information–theoretic relation   
\begin{equation}
\frac{\left\langle\mathcal{W}\right\rangle - \Delta F}{k_{B}T}=\left\langle\mathcal{I}_{x,x_{\varphi}}\right\rangle,
\label{jar1}
\end{equation}
where $\left\langle\mathcal{I}_{x,x_{\varphi}}\right\rangle$ is the mutual information between the read-out distribution for the final probe state (where the parameter $\varphi$ is encoded, $\rho_{\varphi}$) and the distribution for the initial thermal state ($\rho_{T}$). $x$ is some parameter characterizing the distribution. Employing a different approach, Vedral \cite{Vedral} showed that from the averaged exponential of (\ref{jar1}) is possible to obtain the Jarzynski equality (see also the Supplementary Information).   

In addition to the above results, we can show that
\begin{equation}
\left\langle\mathcal{I}_{x,x_{\varphi}}\right\rangle \approx \frac{\delta^{2}_{\varphi}}{2}\mathcal{F},
\label{aux}
\end{equation}
where $\mathcal{F}$ is the Fisher information (classical or quantum) for the encoded state $\rho_{\varphi}$ and
\begin{equation}
\left(\frac{\delta_{\varphi}}{\varphi^{id}}\right)^{2} = \left(\frac{\varphi^{re}}{\varphi^{id}} - 1\right)^{2} \ll 1
\label{approxserror}
\end{equation}
is the relative accuracy of the estimation process. The approximation in Eq. (\ref{approxserror}) means that the error in the measurement must be much smaller than the parameter being measuring. This is quite a reasonable assumption since an error of the same order of the parameter would render meaningless the entire parameter estimation process. 

Combining everything together, from Eqs. (\ref{cramerrao}), (\ref{jar1}) and (\ref{aux}), we show that 
\begin{equation}
(\Delta\varphi)^{2} \geq \frac{\delta_{\varphi}^{2}}{2\langle\mathcal{I}_{x,x_{\varphi}}\rangle},
\end{equation}
which is our main result expressed in Eq. (\ref{main}) as a trade-off relation.

\subsection*{Acknowledgments}
We warmly acknowledge K. Modi for insightful discussions and useful comments. This work was supported by FAPESP, CAPES, CNPq and the Brazilian National Institute for Science and Technology of Quantum Information (INCT-IQ).

\pagebreak

\onecolumngrid
\appendix

\section*{Supplementary Information}

\section{Information-work relation}

In this section we will obtain relation (6) of the main text. Let us consider a quantum system externally driven through some process such that its initial Hamiltonian is described by $H = \sum_n E_n \proj{\psi_n}{\psi_n}$ (in the spectral basis) and its final Hamiltonian reads $H' = \sum_m E'_m \proj{\psi'_m}{\psi'_m}$. The initial system state is taken as $\rho_0$ and the probability distribution for the occupation of the initial Hamiltonian eigenstates is given by $p(n)= \tr \rho_0 \proj{\psi_n}{\psi_n} $. Considering that the system evolves through some process to a final state $\rho'$, we obtain a distribution $p(m)= \tr \rho' \proj{\psi'_m}{\psi'_m} $ for the occupation of the final Hamiltonian eigenstates. A key quantity in our derivation is the mutual information between the joint probability distribution of the outcomes in the measurements of the initial and final Hamiltonians eigenstates, $p(m,n)$. This mutual information can be obtained from the information density $\mathcal{I}_{n,m} \equiv \log\left[p(n,m)/p(n)p(m)\right]$, as
\[
\media{\mathcal{I}_{n,m}} = \sum_{m,n} p(m,n) \log \frac{p(m,n)}{p(m)p(n)}.
\]

In a microscopic thermodynamics description, it is possible to reach the Gibbs ensemble from a distribution which maximizes the Shannon entropy satisfying normalization and thermal energy constrains \cite{Jaynes}. Employing the same reasoning we will find the distribution $p(m,n)$ which provides us the maximum information $\media{I_{n,m}}$, during the process that changes the system Hamiltonian from $H$ to $H'$, with the following constrains:
\begin{eqnarray}
\label{constrain1}
\sum_{m,n} p(m,n) &=& 1;\\ \label{constrain2}
\sum_{m,n} (E'_m - E_n) p(m,n) &=& \media{H'} - \media{H} = \Delta E.
\end{eqnarray}

Assorting Lagrange multipliers $\lambda_0$ and $\lambda_1$ to the first and the second constrains, we have
\begin{equation}
\del{}{p(k,l)} \left[ \sum_{m,n} p(m,n) \log \frac{p(m,n)}{p(m)p(n)} %
+ \lambda_0 \left( \sum_{m,n} p(m,n) - 1 \right) %
+ \lambda_1 \left( \sum_{m,n} (E'_m - E-n) p(m,n) - \Delta E \right) \right] = 0.
\label{variations}
\end{equation}
Since the variations on the probability distribution elements are independent, Eq. (\ref{variations}) is satisfied if 
\begin{equation*}
\log \frac{p(k,l)}{p(k)p(l)} + 1 + \lambda_0 + \lambda_1 (E'_k - E_l) = 0.
\end{equation*}
Therefore $ p(k,l) = p(k)p(l) e^{ - 1 - \lambda_0 - \lambda_1 (E'_k - E_l) }$. From the normalization constrain introduced in Eq. (\ref{constrain1}), it follows that
\begin{equation*}
\sum_{m,n} p(m)p(n) e^{ - 1 - \lambda_0 - \lambda_1 (E'_m - E_n) } %
= e^{ - 1 - \lambda_0 } \sum_{m,n} p(m)p(n) e^{ - \lambda_1 (E'_m - E_n) } = 1,
\end{equation*}
this implies $e^{ 1 + \lambda_0 } = \Zeta$, where we have defined 
$\Zeta \equiv \sum_{m,n} p(m)p(n) e^{ - \lambda_1 (E'_m - E_n) }$. In this way, we can rewrite the joint probability distribution of the initial and final outcomes as
\begin{equation}
\label{prob-z}
p(m,n) = p(m)p(n) \frac{1}{\Zeta} e^{ - \lambda_1 (E'_m - E_n) }.
\end{equation}

The conditional probability for the occurrence of outcome $m$ in a measurement, on $\rho'$, of the final Hamiltonian eigenstates, $H'$, given that the initial outcome was $n$, is $p(m|n) =p(n,m)/p(n)$, which, from the above relations, turns out to be $p(m|n) = e^{ - \lambda_1 (E'_m - E_n) } p(m)/\Zeta$ ($\sum_m p(m|n) = 1$). 

From the energy constrain in Eq. (\ref{constrain2}), we obtain
\begin{align*}
\sum_{m,n} E'_m p(m,n) &= \sum_m E'_m p(m) = \media{H'}, \\
\sum_{m,n} E_n p(m,n) &= \sum_n E_n p(n) = \media{H}.
\end{align*}
We note that $p(n)$ is independent from $\lambda_1$, so $\media{H}$ does not fix the $\lambda_1$ value, being $\lambda_1$ be taken as an arbitrary constant expressed in the inverse of energy unit. For any finite $\lambda_1$ (with $|\lambda_1|< \infty$), we can use Eq. (\ref{prob-z}) to write 
\begin{equation}
\sum_{m,n} (E'_m - E_n) p(m)p(n) \frac{1}{\Zeta} e^{ - \lambda_1 (E'_m - E_n) } = \Delta E.
\end{equation}
It is easy to see that $- \del{\log\Zeta}{\lambda_1} = \Delta E$. Now,let us employ the above relation to rewrite the mutual information between the outcomes in the measurements of the initial and final Hamiltonians eigenstates as
\begin{equation}
\media{I_{n,m}}  = - \log\Zeta - \lambda_1 \Delta E.
\end{equation}
Taking the variation of $\media{I_{n,m}}$ relative to $\Delta E$, we have
\begin{align}
\del{\media{I_{n,m}}}{(\Delta E)} &= -\del{\log \Zeta}{\lambda_1} \del{\lambda_1}{(\Delta E)} - \del{\lambda_1}{(\Delta E)} \Delta E - \lambda_1 \nonumber\\
&= -\lambda_1.
\end{align}
Since $\lambda_1 (E'_m - E_n)$ must be dimensionless and $\media{I_{n,m}}$ is a kind of entropy variation, we can assume $-\lambda_1 = \beta = 1/k_BT$ (where $k_B$ is the Boltzmman constant and $T$ the absolute temperature). Here, we have also considered that the evolution should produce some entropy.

Finally, considering the initial system state as an equilibrium Gibbs state $\rho_0 = e^{-\beta H}/Z$ (with $Z \equiv \tr e^{-\beta H}$ and an unitary transformation driving $H$ to $H'$, we have $p(m|n) = |\bra{\psi'_m} U \ket{\psi_n}|^2$. From Eq. (\ref{prob-z}) we can obtain $p(m)p(n) = \Zeta e^{-\beta(E'_m - E_n)} p(m,n)$. Summing it over the initial and final states, follows that $\sum_{m,n} p(m)p(n) = 1$. So, we can write \cite{Tasaki}  
\begin{align}
\sum_{m,n} p(m)p(n) &= \sum_{m,n} \Zeta e^{-\beta(E'_m - E_n)} p(m,n) \nonumber\\
&= \Zeta \sum_{m,n} e^{-\beta(E'_m - E_n)} p(n)p(m|n) \nonumber\\
&= \Zeta \sum_{m,n} e^{-\beta(E'_m - E_n)} \frac{e^{-\beta E_n}}{Z} |\bra{\psi'_m} U \ket{\psi_n}|^2 \nonumber\\
&= \Zeta \frac{1}{Z} \sum_m e^{-\beta E'_m} \sum_n \bra{\psi'_m} U \ket{\psi_n} \bra{\psi_n} U^\dagger \ket{\psi'_m} \nonumber\\
&= \Zeta \frac{1}{Z} \sum_m e^{-\beta E'_m} \nonumber\\
&= \Zeta \frac{Z'}{Z} = 1
\end{align}
This implies $\Zeta = Z/Z'$, where $Z' \equiv \sum_m e^{-\beta E'_m}$. We note that the final state $\rho'$ is not necessarily an equilibrium state in the above development and the system evolution is also not necessarily adiabatic or energy conserving \cite{Tasaki}. $Z'$ works as partition function for the final Hamitonian $H'$. In fact, this quantity will introduces a connection between equilibrium and non-equilibrium system properties.    

Defining the averaged work, $\langle\mathcal{W}\rangle \equiv \Delta E$ and the Helmholtz Free energy as $F \equiv -k_B T \log Z$, we can write
\begin{align}
\media{\mathcal{I}_{n,m}} &= - \log\frac{Z}{Z'} + \frac{1}{k_B T} \langle\mathcal{W}\rangle \nonumber\\
&= \frac{\langle\mathcal{W}\rangle -  \Delta F }{k_B T} \nonumber\\
&=\beta (\langle\mathcal{W}\rangle -  \Delta F).
\end{align}
This last equation is compatible with the result obtained in Ref. \cite{Vedral} by other methods. If one takes the averaged exponential of the information density, it results in $\media{\exp\{-\mathcal{I}_{n,m}\}}=1$, which implies Jarzinski equality  
\begin{equation}
\media{e^{-\beta \mathcal{W}}} = e^{-\beta \Delta F},
\end{equation} 
This was also showed in Ref. \cite{Vedral} by a different approach.

\section{Dissipation-Information acquisition Inequality}

In this section we back to our general description of a parameter estimation process described in Fig. 1(a) of the main text. Using the results introduced in the previous section, we will obtain an inequality for the acquired information in a parameter estimation process and the work dissipated during the process. In this scenario the mutual information introduced above quantify the correlations between the probe system in the thermal state (before the probe initial preparation) and the after encoding probe second and the first measurements. Let us suppose that during the parameter estimation protocol, the probe system is driven from the thermal $\rho_T$ to an encoded state $\rho_\varphi$ such that the initial Hamiltonian is $H = \sum_n E_n \proj{\psi_n}{\psi_n}$ the final one reads $H' = \sum_m E'_m \proj{\psi'_m}{\psi'_m}$, in a similar way to what was done in the previous section.

Here, we consider two distinct processes, i.e., the ideal (reversible, theoretical) and the real (irreversible, experimental). The difference between then is that the second one includes a small deviation from the reversible dynamics. We are interested in a hight precision parameter estimation device. Such a device should work near to a reversible dynamics in order to obtain hight precision. In that sense, we consider that the actual (real) dynamics of the system is irreversible as a slightly deviation from reversible dynamics. All the development bellow consider this scenario. 

The difference in the mutual information densities between the ideal and the real processes is given by 
\begin{eqnarray}
\mathcal{I}_{0:\varphi}^{re}\left(  j,k\right)  -\mathcal{I}_{0:\varphi}^{id}\left(  j,k\right)   &=& \ln p_{0\varphi}^{re}\left(j,k\right) -\ln\left[p_{0}^{re}\left(j\right)p_{\varphi}^{re}\left(k\right)\right] - \ln p_{0\varphi}^{id}\left(j,k\right) + \ln\left[p_{0}^{id}\left(j\right)p_{\varphi}^{id}\left(k\right)\right] \nonumber \\
&=& \ln\frac{p_{0\varphi}^{re}\left(j,k\right)}{p_{0\varphi}^{id}\left(j,k\right)} - \ln\frac{p_{0}^{re}\left(j\right)}{p_{0}^{id}\left(  j\right)} - \ln\frac{p_{\varphi}^{re}\left(k\right)}{p_{\varphi}^{id}\left(k\right)},
\end{eqnarray}
where the labels $re$ and $id$ means the real and the ideal processes, respectively. Multiplying this last equation by $p_{0\varphi}^{re}\left(j,k\right)$ and summing over $j$ and $k$ we obtain
\begin{eqnarray}
\left\langle\mathcal{I}_{0:\varphi}^{re}\left(j,k\right)\right\rangle_{re} &-& \left\langle\mathcal{I}_{0:\varphi}^{id}\left(j,k\right)\right\rangle _{re} = \sum\limits_{j,k}p_{0\varphi}^{re}\left(j,k\right)  \left[\ln\frac{p_{0\varphi}^{re}\left(j,k\right)}{p_{0\varphi}^{id}\left(j,k\right)} - \ln\frac{p_{0}^{re}\left(j\right)  }{p_{0}^{id}\left(  j\right)  }-\ln\frac{p_{\varphi}^{re}\left(k\right)}{p_{\varphi}^{id}\left(  k\right)  }\right]  \nonumber\\
& =&S(p_{0\varphi}^{re}||p_{0\varphi}^{id})-S(p_{0}^{re}||p_{0}^{id})-S(p_{\varphi}^{re}||p_{\varphi}^{id}).
\label{Ap2}
\end{eqnarray}
In this last expression, $S(p||q)=\sum_{a}p\left(a\right)\ln\left[p\left(a\right) / q\left(a\right)\right]$ is the relative entropy between distributions $p$ and $q$. $\left\langle\mathcal{\cdot}\right\rangle _{re}$ means that the average is taken over the real (non-ideal) process probability distribution.

Using the results of the previous section of this Supplementary Information, we can write the averaged value of the information density as 
\begin{align}
\media{\mathcal{I}_{0\varphi}^{re}\left(  j,k\right)}_{re} &= \beta\left(\langle\mathcal{W}^{re}\rangle - \Delta F^{re}\right) = \frac{\left\langle \mathcal{W}_{D}^{re}\right\rangle }{k_{B}T},
\end{align}
with $\left\langle\mathcal{W}_{D}^{re}\right\rangle$ being the mean dissipated work during the real (non-ideal) implementation of the parameter estimation process. 

In the next section we show that
\begin{equation}
S(p_{0\varphi}^{re}||p_{0\varphi}^{id})\approx\frac{\delta_{\varphi}^{2}}{2}\mathcal{F}\left(p_{\varphi}^{id}\right),
\label{Ap3}
\end{equation}
with $\delta_{\varphi}$ being the accuracy of the implementation of the process and $\mathcal{F}\left(p_{\varphi}^{id}\right)$ the Fisher information of the final ideal distribution (see main text and the next section of this Supplementary Information for the physical significance and precise mathematical definition of these quantities). 

Putting all these results together we obtain
\begin{equation}
\frac{\left\langle \mathcal{W}_{D}^{re}\right\rangle }{k_{B}T}-\frac{\delta_{\varphi}^{2}}{2}\mathcal{F}\left(  p_{\varphi}^{id}\right)=\Omega,\nonumber
\end{equation}
with
\begin{equation}
\Omega=\left\langle \mathcal{I}_{0\varphi}^{id}\left(  j,k\right)\right\rangle _{re}-S(p_{0}^{re}||p_{0}^{id})-S(p_{\varphi}^{re}||p_{\varphi}^{id}).
\end{equation}

As a consequence of the inequality in Cram\'{e}r-Rao relation (Refs. \cite{Cramer,Rao}) our main result, Eq. (1) of the main text, is also an inequality. Therefore, to prove it all we have to do is to prove that $\Omega \geq 0$ for all distributions. Thus
\begin{eqnarray}
\Omega &=& \sum\limits_{j,k} p_{0\varphi}^{re}\left(  j,k\right)\left[\log p_{0\varphi}^{id}\left(j,k\right) - \log p_{0}^{id}\left(j\right) - \log p_{\varphi}^{id}\left(k\right)\right] \nonumber \\
&-&\sum\limits_{j}p_{0}^{re}\left(j\right)\left[\log p_{0}^{re}\left(j\right) - \log p_{0}^{id}\left(j\right)\right] \nonumber \\
&-& \sum\limits_{k}p_{\varphi}^{re}\left(k\right)\left[\log p_{\varphi}^{re}\left(k\right) - \log p_{\varphi}^{id}\left(k\right)  \right] \nonumber \\
&=& \sum\limits_{j,k}p_{0\varphi}^{re}\left(j,k\right)  \log p_{0\varphi}^{id}\left(j,k\right) + H\left(p_{0}^{re}\right) + H\left(p_{\varphi}^{re}\right),
\end{eqnarray}
where $H\left(p\right)  =-\sum_{k}p\left(k\right)\log p\left(k\right)$ is the Shannon entropy. The first term of this expression measures our lack of knowledge about the ideal probability distribution. In fact, $\log p_{0\varphi}^{id}$ is a measure o the information contained in the ideal distribution (the one we expected to happen). However, events occur accordingly with the real distribution $p_{0\varphi}^{re}$ (due to the finite precision of the experimental apparatus). This is the cause of the loss of information. It is not difficult to show that
\begin{equation}
H\left(  p_{0\varphi}^{re}\right)  \leq - \sum\limits_{j,k}p_{0\varphi}^{re}\left(j,k\right)\log p_{0\varphi}^{id}\left(  j,k\right),  
\end{equation}
for every probability distribution. Then, in order to have $\Omega\geq 0$, the following relation must be obeyed
\begin{equation}
H\left(p_{0}^{re}\right) + H\left(p_{\varphi}^{re}\right) \geq - \sum\limits_{j,k}p_{0\varphi}^{re}\left(j,k\right)  \log p_{0\varphi}^{id}\left(j,k\right) \geq  H\left(p_{0\varphi}^{re}\right),
\end{equation}
which is the well know superaditivity relation for the entropy. Thus, we are lead to conclude that
\begin{equation}
\frac{\left\langle\mathcal{W}_{D}^{re}\right\rangle }{k_{B}T} -\frac{\delta_{\varphi}^{2}}{2}\mathcal{F}\left(p_{\varphi}^{id}\right)\geq 0,
\end{equation}
which proves Eq. (9).

\section{Fisher information and relative entropy}

One way to define the Fisher information of a probability distribution $p_{\varphi}$ is through the calculation of the relative entropy between $p_{\varphi}$ and $p_{\varphi+\delta_{\varphi}}$ yielded by a small shift $\delta_{\varphi}$ in the parameter $\varphi$. $\delta_{\varphi}$ is then a measure of the accuracy of the process, i.e., the minimum error compatible with the specific process under consideration. It quantifies how much the real probability distribution diverges from the ideal one. We can write the relative entropy $S(p_{\varphi}\|p_{\varphi+\delta_{\varphi}})$ as (see Refs. \cite{Kullback,Hayashi})
\begin{flalign}
\label{RelEnt}
S(p_{\varphi}\|p_{\varphi+\delta_{\varphi}}) &= \sum_{j,k} p_{\varphi} (j,k) \ln \frac{p_{\varphi} (j,k)}{p_{\varphi+\delta_{\varphi}} (j,k)}\nonumber\\
&= -\sum_{j,k} p_{\varphi} (j,k) \ln \frac{p_{\varphi+\delta_{\varphi}} (j,k)}{p_{\varphi} (j,k)}.
\end{flalign}
By using a Taylor expansion about $\delta_{\varphi}$ we can write
\begin{flalign}
\label{expansion}
\ln p_{\varphi+\delta_{\varphi}} - \ln p_{\varphi} = \delta_{\varphi} \frac{\partial \ln p_{\varphi}}{\partial\varphi} + \frac{\delta^{2}_{\varphi}}{2} \frac{\partial^2 \ln p_{\varphi}}{\partial\varphi^2} + \mathcal{O}(\delta^{3}_{\varphi}).
\end{flalign}

Now, substituting \eqref{expansion} in \eqref{RelEnt}, the first order term yields
\begin{equation}
-\delta \sum_{j,k} \frac{p_{\varphi} (j,k)}{p_{\varphi}(j,k)} \frac{\partial p_{\varphi}(j,k)}{\partial\varphi} = -\delta_{\varphi} \frac{\partial}{\partial\varphi} \sum_{j,k} p_{\varphi} (j,k) = 0
\end{equation}
while the second order one leads us to
\begin{equation}
- \frac{\delta^{2}_{\varphi}}{2} \sum_{j,k} p_{\varphi} (j,k) \frac{\partial^2 \ln p_{\varphi}}{\partial\varphi^2} =
- \frac{\delta^{2}_{\varphi}}{2} \sum_{j,k} \frac{p_{\varphi} (j,k)}{p_{\varphi} (j,k)} \frac{\partial^2 p_{\varphi}(j,k)}{\partial\varphi^2}
 + \frac{\delta^{2}_{\varphi}}{2} \sum_{j,k} p_{\varphi} (j,k) \left( \frac{1}{p_{\varphi} (j,k)} \frac{\partial p_{\varphi}(j,k)}{\partial\varphi} \right)^2
\end{equation}

Since the probability distribution is normalized,
\begin{equation}
\sum_{j,k} \frac{\partial^2  p_{\varphi}}{\partial\varphi^2} = 0,
\end{equation}
and keeping terms up to second order we obtain
\begin{eqnarray}
S(p_{\varphi}\|p_{\varphi+\delta_{\varphi}}) &\approx& \frac{\delta^{2}_{\varphi}}{2} \sum_{j,k} p_{\varphi} (j,k) \left( \frac{\partial}{\partial\varphi} \ln p_{\varphi}(j,k) \right)^2 \nonumber\\
&=& \frac{\delta^{2}_{\varphi}}{2} \mathcal{F}(p_{\varphi}),
\end{eqnarray}
where $\mathcal{F}(p_{\varphi})$ is the Fisher information of $p_{\varphi}$. This approximation implies that we must have
\begin{equation}
\left|\frac{\delta_{\varphi}}{\varphi}\right| \ll 1,
\end{equation}
meaning that the error in the measurement is much smaller than the parameter we are measuring. This is a quite reasonable assumption since an error of the same order of magnitude of the parameter would turn meaningless the measurement process. 

In order to prove Eq. (\ref{Ap3}) let us introduce the {\em Jeffreys' divergence} (Ref. \cite{Kullback} of the manuscript)
\begin{eqnarray}
J(p_{\varphi},p_{\varphi+\delta_{\varphi}}) &=& S(p_{\varphi}\|p_{\varphi+\delta_{\varphi}}) + S(p_{\varphi+\delta_{\varphi}}\|p_{\varphi}) \nonumber\\
&=& \sum_{j,k} \left( p_{\varphi} (j,k) - p_{\varphi+\delta_{\varphi}} (j,k) \right) \ln \frac{p_{\varphi} (j,k)}{p_{\varphi+\delta_{\varphi}} (j,k)}.
\label{eqjef1}
\end{eqnarray}
Since (\ref{eqjef1}) is symmetrical, we may rewrite it as 
\begin{equation}
J(p_{\varphi},p_{\varphi+\delta_{\varphi}}) = \sum_{j,k} \Delta p_{\varphi+\delta_{\varphi}}(j,k) \ln \left( 1 + \frac{\Delta p_{\varphi+\delta_{\varphi}}(j,k)}{p_{\varphi} (j,k)} \right)
\end{equation}
where $\Delta p_{\varphi+\delta_{\varphi}} = p_{\varphi+\delta_{\varphi}} - p_{\varphi}$. Due to the fact that $\delta_{\varphi}$ is small, the distribution $p_{\varphi+\delta_{\varphi}}$ will be close to $p_{\varphi}$. Therefore
\begin{equation}
\ln \left( 1 + \frac{\Delta p_{\varphi+\delta_{\varphi}}(j,k)}{p_{\varphi} (j,k)} \right) \approx \frac{\Delta p_{\varphi+\delta_{\varphi}}(j,k)}{p_{\varphi} (j,k)}
\end{equation}
allowing us to write
\begin{flalign}
J(p_{\varphi},p_{\varphi+\delta_{\varphi}}) &\approx \sum_{j,k} p_{\varphi} (j,k) \left( \frac{\Delta p_{\varphi+\delta_{\varphi}}(j,k)}{p_{\varphi} (j,k)} \right)^2 \nonumber\\
&= \sum_{j,k} p_{\varphi} (j,k) \left( \frac{\delta_{\varphi}}{p_{\varphi}} \frac{\Delta p_{\varphi+\delta_{\varphi}}(j,k)}{\delta_{\varphi}} \right)^2 \nonumber\\
&= \delta^{2}_{\varphi} \sum_{j,k} p_{\varphi} (j,k) \left( \frac{1}{p_{\varphi} (j,k)} \frac{\partial p_{\varphi}(j,k)}{\partial\varphi}   \right)^2 \nonumber\\
&= \delta^{2}_{\varphi} \mathcal{F}(p_{\varphi})
\end{flalign}

From equation (\ref{eqjef1}) we have $S(p_{\varphi+\delta_{\varphi}}\|p_{\varphi}) = J(p_{\varphi},p_{\varphi+\delta_{\varphi}}) -  S(p_{\varphi}\|p_{\varphi+\delta_{\varphi}})$, so
\begin{equation}
S(p_{\varphi+\delta_{\varphi}}\|p_{\varphi}) \approx \frac{\delta^{2}_{\varphi}}{2} \mathcal{F}(p_{\varphi})
\end{equation}

Finally, making the identifications $p_{\varphi} \equiv p_{0\varphi}^{id}$ and $p_{\varphi+\delta_{\varphi}} \equiv p_{0\varphi}^{re}$ we obtain Eq. (\ref{Ap3}), thus proving our main claim, Eq. (1) of the manuscript.


\end{document}